\documentclass[usenatbib,usegraphicx]{mn2e}
\usepackage{subfigure}
\usepackage{longtable}
\usepackage{graphicx}
\usepackage{graphics}
\usepackage{keyval}
\usepackage{trig}
\usepackage{rotating}
\usepackage{epstopdf}
\def\beq{\begin{equation}}
\def\eeq{\end{equation}}
\def\bey{\begin{eqnarray}}
\def\eey{\end{eqnarray}}

\def\sun{\odot}
\def\lsim{\mathrel{\raise.3ex\hbox{$<$\kern-.75em\lower1ex\hbox{$\sim$}}}}
\def\gsim{\mathrel{\raise.3ex\hbox{$>$\kern-.75em\lower1ex\hbox{$\sim$}}}}

\def \farcs{\hbox{$.\!\!^{\prime\prime}$}}

\title[Weak lensing by galaxies: implications for MOND] {The relation
between stellar mass and weak lensing signal around galaxies:
Implications for MOND}

\author[Lanlan Tian, Henk Hoekstra, Hongsheng Zhao]
{Lanlan. Tian$^{1}$\thanks{email:ltian@uvic.ca}, 
Henk Hoekstra$^{1}$\thanks{Alfred P. Sloan fellow}, 
Hongsheng Zhao$^{2}$\\   \\
$^{1}$Department of Physics and Astronomy, University of Victoria,
Victoria, BC, V8W 2Y2, Canada\\
$^{2}$SUPA, School of Physics and Astronomy, University of St. 
Andrews, KY16 9SS Scotland\\
}

\begin{document}

\date{Accepted ... Received ... ; in original form ...}

\pagerange{\pageref{firstpage}--\pageref{lastpage}} \pubyear{2008}

\maketitle

\label{firstpage}

\begin{abstract}

We study the amplitude of the weak gravitational lensing signal as a
function of stellar mass around a sample of relatively isolated
galaxies. This selection of lenses simplifies the interpretation of
the observations, which consist of data from the Red-sequence Cluster
Survey and the Sloan Digital Sky Survey. We find that the amplitude of
the lensing signal as a function of stellar mass is well described by
a power law with a best fit slope $\alpha=0.74\pm0.08$. This result is
inconsistent with Modified Newtonian Dynamics, which predicts
$\alpha=0.5$ (we find $\alpha>0.5$ with 99.7\% confidence). As a
related test, we determine the MOND mass-to-light ratio as a function
of luminosity. Our results require dark matter for the most luminous
galaxies ($L \gsim 10^{11}L_\odot$). We rule out an extended halo of gas or
active neutrinos as a way of reconciling our findings with MOND. Although we
focus on a single alternative gravity model, we note that our results
provide an important test for any alternative theory of gravity.

\end{abstract}


\section{Introduction}
\protect\label{sec:intr}

It is now well established that there are significant discrepancies
between the Newtonian gravitational mass and the observable luminous
mass on scales ranging from galaxies to clusters of galaxies. Two
fundamentally different explanations have been proposed to solve these
observations. The current paradigm is that General Relativity (GR) is
correct on large scales, but that the derived gravitational mass is
larger because of large amounts of unseen matter.  This has led to the
current standard cosmological model of a universe dominated by Cold
Dark Matter and a cosmological constant ($\Lambda$CDM). This model is
well-developed and widely applied to galactic and cosmological
observations.  It fits a range of observations, most notably the
cosmic background radiation, extremely well (e.g., Spergel et
al. 2007). The need for dark matter to explain astronomical
observations has been a long-standing issue and a number of dark
matter candidates, inspired by particle physics, have been suggested.

The current lack of a direct detection of the dark matter particle has
led to an alternative approach to explain the observations. Instead of
invoking dark matter, it is assumed that the law of gravity differs
from the Newtonian gravity on a scale of weak gravity which cannot yet
be reproduced in current gravitational experiments.  One of the most
studied alternatives is Modified Newtonian Dynamics (MOND) proposed by
Milgrom (1983a). It has evolved over the past 25 years from an
empirical fit to galaxy rotation curve data (Milgrom 1983b, Sanders \&
McGaugh 2002) to a relativistic tensor-vector-scalar theory (TeVeS,
Bekenstein 2004). Recent developments include a theory of a vector
field with a non-linear coupling to the space-time metric (Zlosnik,
Ferreira, Starkman 2007, Zhao 2007).

MOND works particularly well on galactic scales (for a review, see
Sanders \& McGaugh 2002) and this is where most (dynamical) tests have
focused on (e.g. Read \& Moore 2005, Famaey et al 2006, Nipoti et al
2007, Corbelli \& Salucci 2006, Gentile et al. 2007, Zhao \& Famaey
2006, Famaey \& Binney 2005). There are a few tests on sub-galactic
scales using the velocity dispersion of globular clusters (Baumgardt et
al. 2005) and using the tidal radius (Zhao 2005, Zhao \& Tian 2006).

Our approach differs from previous studies in a number of ways. Unlike
GR, MOND is a non-linear gravity theory, resulting in fundamental
differences in their global scaling relations. Hence, rather than
comparing the strength of the gravitational potential, we focus on how
it changes with (baryonic) mass, although we do consider an example of
the former as well. A fundamental property of MOND is its `prediction'
of a Tully-Fisher relation (Tully \& Fisher 1977). In MOND this
relation follows from the theory (Milgrom 1983b), whereas in
$\Lambda$CDM it arises from the interplay between dark and baryonic
matter. Also note that observationally the Tully-Fisher relation is
one between the luminosity and the (maximum) rotation velocity, and
thus is a test on sub-galactic scales.

It is therefore useful to examine MOND on scales much larger than
those probed by rotation curves. Probing the gravitational potential
in these outer regions of galaxies provides an ideal test of
alternative gravity, because of the absence of luminous matter (except
for a few satellites and globular clusters). It is in these regions
where MOND and $\Lambda$CDM differ most markedly. In the dark matter
paradigm we would call these regions `dark matter dominated' and in
MOND we call them `deep-MOND regions'.  Finally, rather than dynamics,
we will study the weak gravitational lensing signal around galaxies.
We refer the interested reader to Hoekstra \& Jain (2008) for a recent
review on weak lensing.

We note that other tests involving (strong) gravitational lensing
already have provided important constraints, albeit on relatively
small scales. These studies typically reveal a factor of two
discrepancy between stellar mass and the lensing mass using MOND
(Zhao, Bacon, Taylor, Horne 2006, Chen \& Zhao 2006, Angus et
al. 2007a, Takahashi \& Chiba 2007, Ferreras et al.  2007, Natarajan
\& Zhao 2008).  Of particular intestest is the weak gravitational
lensing study by Clowe et al. (2006) of the `Bullet' cluster. Clowe et
al. (2006) found a large offset between the matter distribution, as
inferred from the lensing analysis, and the gas (which contains most
of of the baryonic mass).  The lensing map, however, does agree well
with the distribution of galaxies, which is expected if both the dark
matter and stars in galaxies are collisionless.

To probe the outer regions around galaxies we employ a technique
called weak gravitational galaxy-galaxy lensing (hereafter g-g
lensing), which is the statistical study of the deformation of distant
galaxies by foreground galaxies. Since the gravitational distortions
induced by an individual lens are too small to be detected, one has to
resort to the study of the ensemble averaged signal around a large
number of lenses. Of particular interest is that the g-g lensing
signal can be measured out to large projected distance, where
dynamical methods are of limited use due to the lack of luminous
tracers. Hence, g-g lensing provides a unique and powerful tool to
probe the gravitational potential on large scales. Only studies of
satellite galaxies can also probe these regions (e.g., Zaritsky \&
White 1994; McKay et al. 2002; Prada et al. 2003; Conroy et al. 2007).

Since the first detection by Brainerd et al. (1996), the accuracy of
g-g lensing studies has improved dramatically thanks to improved
analysis techniques and large amounts of wide-field imaging data
(e.g. Fischer et al.  2000, Hoekstra et al. 2004, Hoekstra et
al. 2005, Mandelbaum et al. 2006, Parker et al. 2007). We refer to
these papers for a more in-depth discussion of this area of research.
Relevant for our study is the availability of (photometric or
spectroscopic) redshift information for the lens galaxies. Only
recently has this kind of information become available for large
samples (Hoekstra et al. 2005, Mandelbaum et al. 2006). As a result we
can now compare how the strength of the g-g lensing signal depends on
the baryonic content of the lenses.

This paper is organized as follows. In \S2, we discuss the expected
dependence of the lensing signal on stellar mass. In \S3 we describe the
galaxy-galaxy lensing data used in our analysis. In \S4 we present our
results and discuss the implications for MOND.  Throughout this paper,
we adopt a Hubble parameter $H_0=70$ km/s/Mpc and all the
error bars correspond to 68\% confidence limits ($1\sigma$).

\begin{figure*}
\def\subfigtopskip{0pt} 
\def\subfigbottomskip{4pt}
\def\subfigcapskip{1pt}
\centering
\begin{tabular}{c}
\subfigure{\label{fig:corfith1}
\includegraphics[angle=0,width=14.5cm]{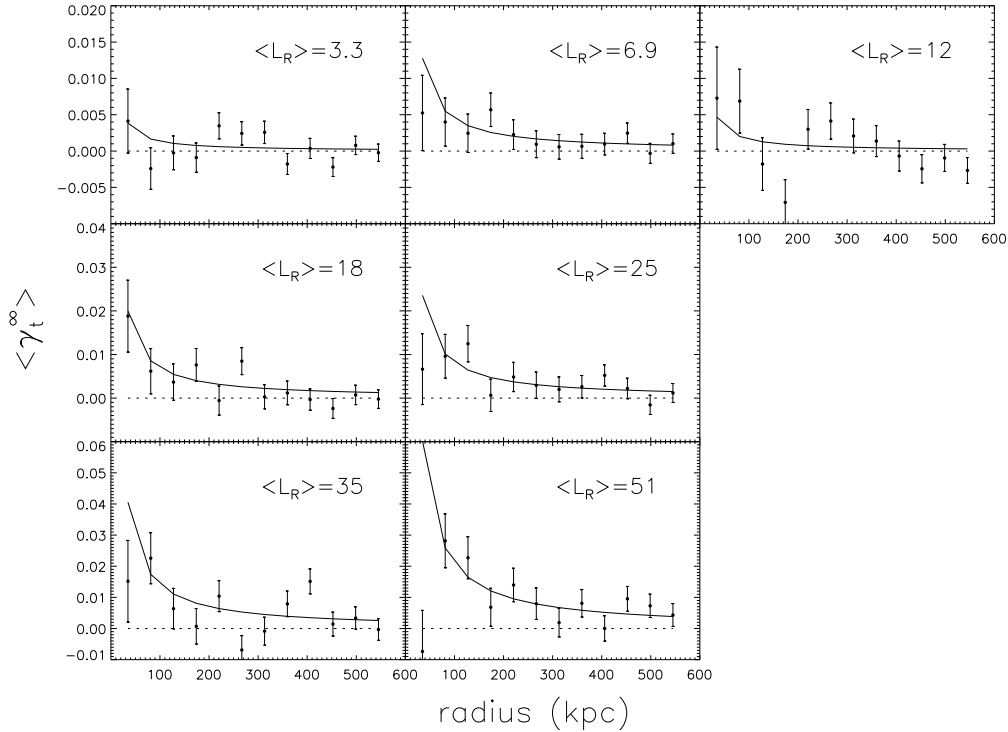}}
\end{tabular}
\caption{The observed ensemble averaged tangential shear around
`isolated' galaxies from Hoekstra et al. (2005). The data are shown
for 7 luminosity bins (with the mean $L_R$ indicated in units of $10^9
L_{R,\sun}$). The solid line indicates the best fit MOND point mass
model. The lensing signal has been scaled to that of a lens at the
average lens redshift ($z \sim 0.32$) and a source redshift of
infinity.}
\label{fig:corfit}
\end{figure*}

\section{Theoretical Predictions}

One of the reasons for the success of MOND is the ability to provide
excellent fits to rotation curves over a wide range in mass, thanks to
its `built-in' Tully-Fisher relation (Milgrom 1983b). As we show
below, this feature has consequences for the predicted scaling of the
lensing signal with stellar mass.

In MOND the only source of gravity is the luminous matter. As we are
concerned with the lensing signal on large scales, we assume that the
galaxy (stellar) mass distribution can be approximated by a point mass
model. We have verified that the lensing signal on large scales ($>20$
kpc) is insensitive to the actual baryonic density profile. This is
expected because the shear at large radii depends on the enclosed
mass, which quickly converges to the same value for sufficiently
compact mass distributions. Under these assumptions the MOND effective
density $\rho_{\rm eff}$ for a point mass with mass $M$ is given by
(see the Appendix for a discussion of the calculation for spherically
symmetric mass distributions):
\begin{equation}
\rho_{\rm eff}(r)=\frac{\nabla^2 \Phi}{4\pi G}=\frac{v_0^2}{4\pi
	G}\frac{1}{r^2} ,
\end{equation}
\noindent where $\Phi$ is the gravitational potential in MOND, $v_0
\equiv {(GMa_0)}^{\frac{1}{4}}$ and $a_0$ is the MOND critical
acceleration ($a_0 \approx 10^{-8}$cms$^{-2}$). We assumed that $r\gg
r_0\equiv \sqrt{GM/a_0}$. For a mass $M=10^{11} M_{\sun}$, we find
$r_0 \approx 10$kpc, which is much smaller than the scales we probe in
this paper. Once we have obtained the effective density $\rho_{\rm
eff}$, we can apply the same procedure as in the GR case to calculate
the tangential shear (Zhao, 2006).  The effective surface density is
given by
\beq
\label{eqn:surf}
\Sigma(r)=\frac{v_0^2}{4G} \frac{1}{r} .
\eeq 
\noindent The convergence $\kappa$ is the ratio of the surface density
and the critical surface density $\Sigma_{\rm crit}$, which is given by
\beq 
\Sigma_{\rm crit}=\frac{c^2}{4\pi G}\frac{D_s}{D_lD_{ls}},
\eeq 
\noindent where $D_{ls}$ is the distance from the lens to the source
and $D_l$ and $D_s$ are the distances from the observer to the lens
and the source, respectively.  We use the fact that the azimuthally
averaged tangential shear $\gamma_t$ (which corresponds to the
observed g-g lensing signal) is related to the convergence through
\(\gamma_t={\bar{\kappa}(<r)} - \kappa(r)\), where $\bar\kappa(<r)$ is
the mean convergence within the radius $r$. This yields a convergence
$\kappa$ and tangential shear $\gamma_t$ given by

\begin{equation}
\label{eqn:mga}
\kappa(r)=\gamma_t(r)=\frac{\Sigma(r)}{\Sigma_{\rm crit}}=\frac{r_E}{2r},
\end{equation} 

\noindent where the Einstein radius $r_E$ is given by

\beq
\label{eqn:mre}
r_E = 2\pi \left(\frac{v_0}{c}\right)^2 \frac{D_{ls}}{D_s}.
\eeq

\noindent Such a tangential shear profile is a good description of the
galaxy-galaxy lensing signal within $\sim 400$kpc (e.g., Hoekstra et
al. 2004). In the GR case, this shear profile corresponds to a
singular isothermal sphere (SIS) with a line-of-sight velocity
dispersion $\sigma_v$ and an Einstein radius given by

\beq
\label{eqn:nre}
r_E=4\pi \left(\frac{\sigma_v}{c}\right)^2\frac{D_{ls}}{D_s},
\eeq 

Although similar in appearance, the physical interpretations are
markedly different.  This becomes apparent when we consider the
dependence of the Einstein radius on mass. In the GR case we have

\beq
\label{eqn:nlre}
r_E \propto \sigma_v^2 \propto M.
\eeq

\noindent Hence, we expect a linear relation between the total galaxy mass
(within a fixed radius) and Einstein radius $r_E$. In the MOND case,
however, we have $v_0^2 \propto \sqrt{M}$ (as $v_0 \equiv
{(GMa_0)}^{\frac{1}{4}}$), which yields

\beq
\label{eqn:mlre}
r_E \propto v_0^2 \propto \sqrt{M_*},
\eeq

\noindent where we explicitly use the stellar mass $M_*$ as the mass
of a galaxy (we ignore the contribution from gas). Therefore MOND
predicts a Tully-Fisher-like scaling relation between the Einstein
radius and the stellar mass. In the GR case, the total mass depends on
the relative contributions of dark and luminous matter, thus
preventing us from predicting the value of the slope.

\begin{figure*}
\def\subfigtopskip{0pt} 
\def\subfigbottomskip{4pt}
\def\subfigcapskip{1pt}
\centering
\begin{tabular}{c}
\subfigure{\label{fig:corfith2}
\includegraphics[angle=0,width=14.5cm]{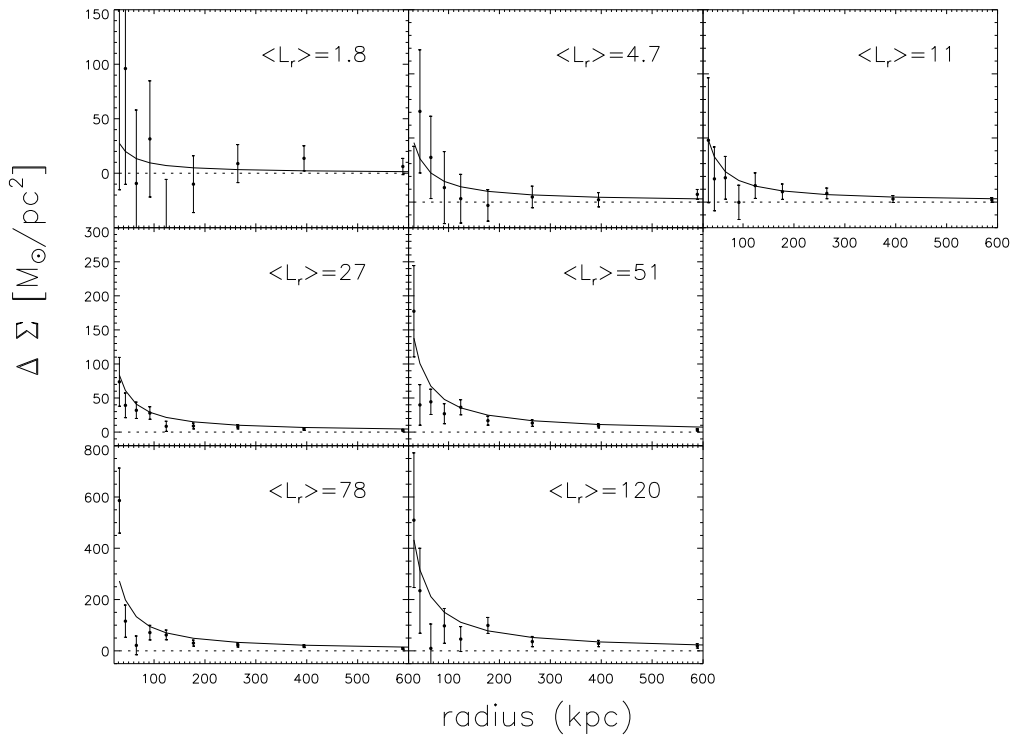}
}
\end{tabular}
\caption{The observed galaxy-galaxy lensing signal around early type
galaxies in low density regions from Mandelbaum et al. (2006). The
data are shown for 7 luminosity bins (with the mean $L_r$ indicated in
units of $10^9 L_{r,\sun}$). The solid line indicates the best fit MOND
point mass model. In order to extract the signal dominated by the lens galaxy
itself, we fit the signals only within $\sim 200$ kpc from the lens.}
\label{fig:SDSSfit} 
\end{figure*}

\section{observational data}
\protect\label{sec:data}

The measurement of the g-g lensing signal as a function of stellar
mass requires a large data set of sources and lenses with redshift
information. As a further complication, the predictions given in \S2
are only valid for an isolated galaxy. If the lensing signal includes
a significant contribution from nearby galaxies, galaxy groups or
clusters, then the inferred Einstein radius will be biased. This is
particularly relevant for faint galaxies (see Fig 7. in Hoekstra et
al. 2005). To ensure that the observed lensing signal is that of the
lens galaxy itself, we consider two particular data sets which are
described in more detail below.

\subsection{Red-sequence Cluster Survey (RCS)}

The Red-sequence Cluster survey (RCS) is a galaxy cluster survey using
$R_c$ and $z'$ imaging data (Gladders \& Yee 2005). Within the
surveyed area, $\sim33.6$ deg$^2$ were also imaged in the $B$ and $V$
bands. The four-filter data in the latter area were used by Hsieh et
al. (2005) to derive photometric redshifts for $1.2\times10^6$
galaxies, which were used by Hoekstra et al. (2005; H05) to study the
weak lensing signal as a function of galaxy properties.  H05 selected
a sample of `isolated' lens galaxies by ensuring that no galaxy more
luminous than the lens was located within 30". Hence the galaxies in
the faintest bin are truly isolated, whereas the brightest galaxies
can have nearby (faint) companions.

The `isolated' lens sample comprises of 94,509 galaxies with
$0.2<z<0.4$ and restframe $R_C$ luminosities. For the analysis we
limit the measurement of the lensing signal to within 600 kpc from the
lens. Figure~\ref{fig:corfit} shows the observed tangential shear
profiles and the best fit MOND point mass model. As discussed in
Benjamin et al.  (2007) the mean source redshift used in H05 was
biased low because of the lack of a reliable training set at high
redshift. Consequently the masses listed in H05 were biased high by
$\sim 15\%$ compared to the results used here. The luminosity shown in
the plots is the mean rest-frame $R$ band luminosity in units of
$10^9L_{R,\sun}$. Finally, random errors in the photometric redshift
estimates of the lenses lead to an underestimate of the true value of
$r_E$, which we correct for by multiplying the observed $r_E$ with a
correction factor determined from mock catalogs as described in H05.

\begin{figure*}
\def\subfigtopskip{0pt} 
\def\subfigbottomskip{4pt}
\def\subfigcapskip{1pt}
\centering
\begin{tabular}{c}
\includegraphics[angle=0,width=16.0cm]{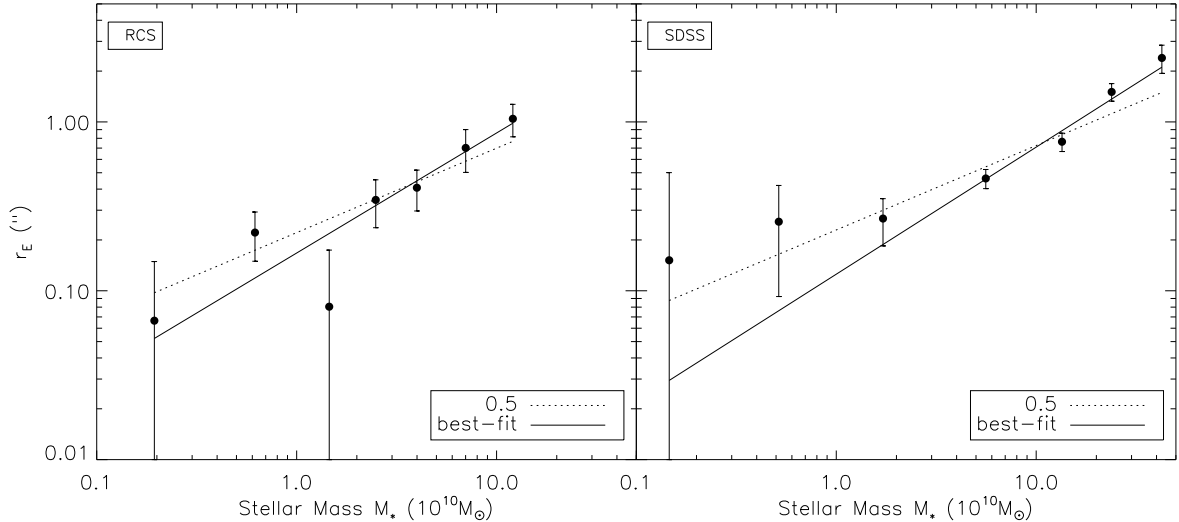}
\end{tabular}
\caption{{\it Left panel:} Einstein radius $r_E$ as a function of
stellar mass (derived from multi-colour photometry) for the RCS data from
Hoekstra et al. (2005). {\it Right panel:} Value for $r_E$ obtained
from the SDSS g-g lensing signal from Mandelbaum et al (2006). To
allow for a simple comparison, all Einstein radii in the plot have
been scaled such that $D_{ls}/D_s=1$.  The dotted line in each plot
represents the best fit assuming $r_E\propto\sqrt{M_*}$ (as predicted
by MOND). The best fit power law is indicated by the solid line.}
\label{fig:rel}
\end{figure*}

\subsection{Sloan Digital Sky Survey (SDSS)}

Mandelbaum et al. (2006; M06) studied the g-g lensing signal using
data from the SDSS survey (York et al. 2000), with the lenses selected
from the SDSS Data Release Four main spectroscopic sample (DR4;
Adelman-McCarthy 2006) which covers 4783 deg$^2$. The lens galaxies
have spectroscopic redshifts between $0.02 < z <0.35$. M06 split their
sample into early and late type galaxies, based on morphology. The
early-type galaxies are also divided into overdense and underdense
samples based on the median local galaxy environment density within
each luminosity bin. As discussed in M06, the local density for each
lens is determined using the spectroscopic galaxy counts in cylinders
of radius 1$h^{-1}$Mpc and a line-of-sight length $\Delta v=\pm
1200$km/s. We use the results for the low-density sample (Fig. 3,
black triangles in M06), because we expect the contribution of
neighboring galaxies to the g-g lensing signal to be reduced. Limiting
the sample to early type galaxies also reduces the variation of
stellar mass-to-light ratio with luminosity.

M06 represent the lensing signal by $\Delta
\Sigma (r)$, where $\Sigma (r)$ is the projected surface density:
\begin{equation}
\Delta \Sigma (r) \equiv \bar{\Sigma}(<r) -
\Sigma(r)=\langle\gamma_t\rangle\Sigma_{\rm crit}.
\end{equation} 
The resulting tangential shear profiles are presented in
Figure~\ref{fig:SDSSfit}. Although the lenses are selected to be in
underdense environments, it is possible that lens galaxies in low
luminosity bins (e.g. L1, L2 and L3) are surrounded by luminous
galaxies. Hence the lensing signals in these low luminosity bins could
include a non-negligible contribution from the surrounding brighter
galaxies.  Theoretical analysis of the expected g-g lensing signal
suggests the group and cluster haloes can dominate the lensing signal
on scales larger than $300$ kpc (Seljak 2000). The signal on scales
less than $\sim 200$ kpc is expected to be dominated by the lens
itself. Therefore we fit a MOND point-mass model only to the
measurements within 200 kpc. The best fit models are represented by
the solid lines in Figure~\ref{fig:SDSSfit}. Note that despite our
concerns, the best fit model is an excellent fit to the points at
large radii and extending the fits to larger radii does not change our
results.

\subsection{Stellar masses}

M06 also present estimates for the stellar
masses. The procedure used by M06 is based on the
same techniques as in Kauffmann et al. (2003) and the stellar masses
are derived from a comparison of a library of star formation history
models to the spectroscopic data. We use the observed (power law)
relation between stellar mass and luminosity to convert the
luminosities listed in Figure~\ref{fig:SDSSfit}. The derived stellar
mass depends predominantly on the adopted low-mass end of the initial
mass function. This leads to an uncertain normalisation, but the
inferred dependence of stellar mass with luminosity is robust (e.g.,
Bell \& de Jong 2001).

Unlike the SDSS data, the RCS results lack a detailed estimate of
stellar masses. However, in addition to numbers for early type
galaxies, M06 also provide stellar masses for
late type galaxies and list the fraction of late types as a function
of luminosity. We use these results to compute the stellar mass as a
function of luminosity for the RCS2 data. H05 computed stellar
mass-to-light ratios as a function of color using the results from
Bell \& de Jong (2001). We compared these (less accurate) results to
our estimates based on the numbers provided in M06 and find good agreement.

\begin{figure}
\includegraphics[angle=0,width=8.5cm]{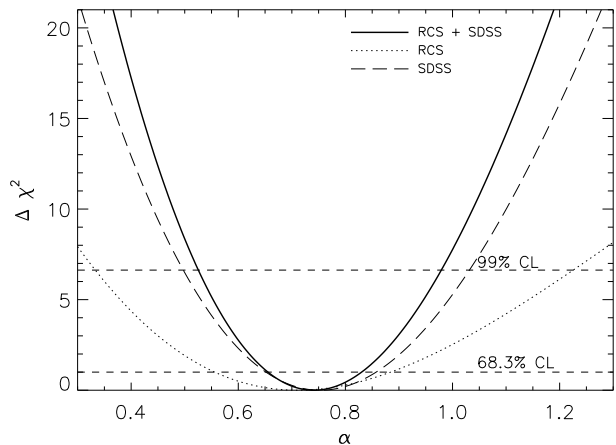}
\caption{$\Delta \chi^2$ as a function of the exponent $\alpha$ in the
power-law relation $r_E \propto M_*^{\alpha}$, while marginalizing
over the normalization. The dotted line corresonds to the constraints
from a fit to the RCS results, whereas the long-dashed line is the
result from the SDSS data. The solid line is the combined
constraint. These results indicate that $\alpha>0.5$ (with 
99.7\% confidence) and thus inconsistent with the MOND
prediction.}\label{fig:chi}
\end{figure}

\begin{figure*}
\def\subfigtopskip{0pt} 
\def\subfigbottomskip{4pt}
\def\subfigcapskip{1pt}
\centering
\begin{tabular}{c}
\includegraphics[angle=0,width=17.0cm]{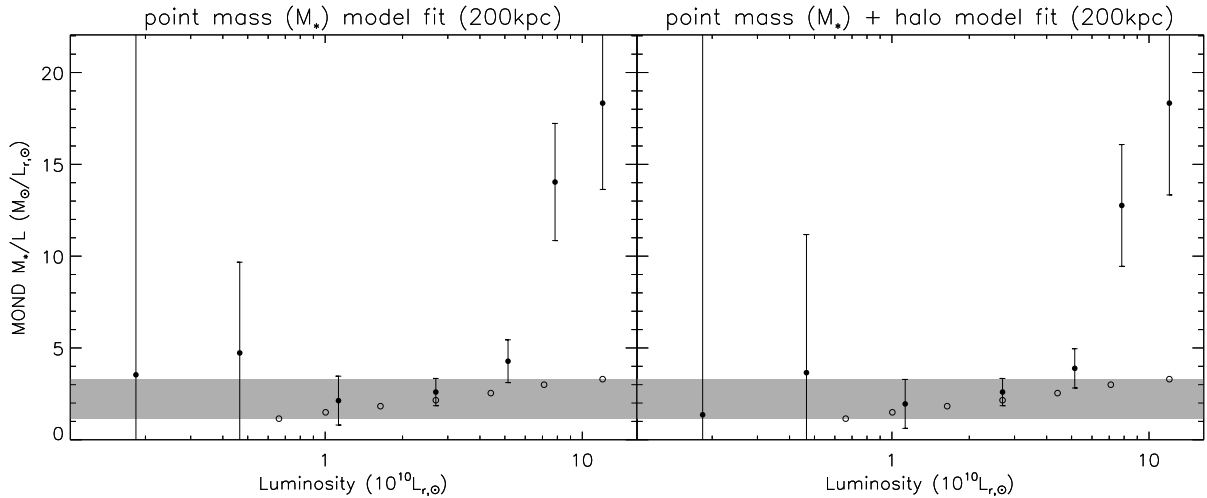}
\end{tabular}
\caption{{\it Left panel:} the inferred MOND mass-to-light ratio as a
function of luminosity.  The derived MOND masses are obtained by
fitting a point mass model to the SDSS lensing data within
200 kpc. Because there is no dark matter in MOND, we take the
derived MOND mass to be the total stellar mass $M_*$ (we can ignore
the contribution from HI).  {\it Right panel:} The MOND mass-to-light
ratio from a fit to the SDSS data when we add a neutrino halo to the
stellar mass. The neutrino halo is assumed to have a $\beta$ profile
and its total mass is three times of the stellar mass. The stellar
mass-to-light ratios as a function of luminosity from Mandelbaum et al
(2006) are indicated by the open circles. The shaded area indicates
the range in those inferred stellar mass-to-light ratios.}
\label{fig:m2l}
\end{figure*}

\section{results}
\protect\label{sec:result}

Figure~\ref{fig:rel} shows the measurement of the Einstein radius as a
function of the stellar mass of the lens. The left panel shows the
results for the RCS data and the right panel corresponds to the
results for the SDSS data. The Einstein radii in Fig.~\ref{fig:rel}
have been scaled to $D_{ls}/D_s=1$. We assume a power-law relation
between the Einstein radius and the stellar mass: $r_E\propto
M_*^\alpha$. For reference, the dotted lines in Figure~\ref{fig:rel}
show the best fit relation for $\alpha=0.5$, which is the expected
slope for MOND.

Before we proceed with our determination of the scaling relation
between lensing signal and stellar mass, we first examine whether the
amplitudes of the signal agree between RCS and SDSS. For the
comparison we adopt $\alpha=0.75$ and obtain a value of
$r_E=0\farcs52\pm0\farcs06$ for a galaxy with a stellar mass of
$M_*=5\times 10^{10}M_\odot$ for the RCS measurements.  For the SDSS
data we obtain a value of $r_E=0\farcs42\pm0\farcs03$, in fair
agreement with the RCS results.

For the RCS data we find a best fit power-law slope of $\alpha=0.71\pm
0.15$ ($\chi_{\rm min}^2=4.5$ for 5 degrees of freedom; we marginalize
over the normalization).  The difference in $\chi^2$
$(\Delta\chi^2=\chi^2-\chi^2_{\rm min})$ as a function of $\alpha$ is
indicated by the dotted line in Figure~\ref{fig:chi}. For the SDSS
data we find $\alpha=0.75\pm 0.09$ ($\chi_{\rm min}^2=4.9$ for 5
degrees of freedom) and the corresponding $\Delta\chi^2$ is indicated
by the long-dashed curve in Figure~\ref{fig:chi}. Combining the RCS
and SDSS constraints yields a value $\alpha=0.74\pm0.08$.  (solid line
in Figure~\ref{fig:chi}). Hence we find good agreement between the RCS
and SDSS data, with the SDSS data providing the best contraint.
Furthermore, the inferred slope for the combined data is larger than
$\alpha=0.5$ (which is the value predicted by MOND) with 99.7\%
confidence. Our results are also in good agreement with Conroy et
al. (2007) who found that the velocity dispersion of satellite
galaxies scales with the stellar mass of the host galaxy
$\sigma\propto M_*^{0.4\pm0.1}$, which corresponds to $r_E\propto
M_*^{0.8}$ for an isothermal sphere model.

An alternative way to present our measurements is to consider the
mass-to-light ratio as a function of luminosity. We expect the
inferred MOND mass-to-light ratio to correspond to the stellar
mass-to-light ratio, because in this case the stellar mass is the only
source of gravity (we can ignore the contribution from neutral
hydrogen). The left panel in Figure~\ref{fig:m2l} shows the derived MOND
mass-to-light ratio as a function of luminosity. The mass is
determined from a fit to the SDSS lensing signal out to
200 kpc, assuming a point mass model for the galaxy.

Interestingly, at low luminosities the mass-to-light ratio is constant
with values in agreement with what one might expect for an old
population (the shaded region indicates the range of stellar M/L from
Mandelbaum et al., 2006). At high luminosities, however, we observe a
significant increase in $M/L$ which is inconsistent with the expected
values for the stellar populations. Hence, these measurements suggest
the need for additional (dark) matter.

\subsection{Potential biases}

The data points at high stellar mass in Figure \ref{fig:rel} (and at
high luminosity in Fig. \ref{fig:m2l} as well) carry much of the
statistical weight, both for RCS and SDSS. It is apparent that if we
were to remove or lower those measurements, the remaining data suggest
a slope consistent with the MOND prediction, instead of the steeper
relation determined in the previous section. Here we argue that those
data points cannot be ignored, because they are in fact the more
reliable measurements.

Compared to brighter galaxies, the lensing signal around a faint
galaxy is more easily affected by the contribution of neighboring
bright galaxies. Although we have selected `isolated' or underdense
lens samples for our analysis, the lensing signal in the low
luminosity bins is potentially biased to higher values, thus lowering
the inferred slope. In contrast, this will not be significant for the
high luminosity bins, especially on small scales. To investigate this
further we examined the effect of environment on the lensing signal
around bright galaxies. The panels for L5b and L6f in Figure 3 from
Mandelbaum et al (2006) correspond to our two highest luminosity
bins. These results suggest there are no significant differences in
the lensing signals on small scales ($\sim 200$ kpc) between the
overdense and underdense samples. This implies that on these small
scales the contribution of the environment to the lensing signals is
negligible and the signal is dominated by the lens itself. Finally,
the agreement between the RCS and SDSS results, which are based on
different selection algorithms for the lenses, suggest the
measurements are robust.

\subsection{Neutrino Halos}

The arguments presented in the previous section suggest the need for
dark matter in luminous galaxies, even in the context of MOND.
Similarly, MOND cannot explain observations of clusters of galaxies
without invoking a significant dark matter component (Sanders, 2007):
the MOND dynamical mass is about a factor $3-4$ times larger than the
observed baryonic mass. Sanders (2007) suggests that active neutrinos
with a mass $m_\nu\sim 2$eV might be able to reconcile these
observations with MOND.

Neutrinos with masses in the range of a few eV cannot accumulate into
dense halos around galaxies because of phase-space constraints
(Tremaine \& Gunn 1979). Nonetheless, as argued by Sanders (2007)
galaxies may be able to acquire a neutrino halo, which would have a
mass $\sim 1.4 m_\nu$ times larger than the baryonic mass. This ratio
of masses is set by the cosmological density of neutrinos and the
baryon density inferred from CMB observations (Spergel et al.,
2007). In this section we examine the effect of such a massive
neutrino halo on the observed g-g lensing signal. Following Sanders
(2007) we consider a mass of $m_\nu=2$eV. We note that this is the
maximum mass allowed by $\beta$-decay experiments, which restrict the
mass of the electron neutrino to a value less than 2eV (Yao et al.,
2006).

We add a neutrino halo model to the stellar component (which we model
by a point mass). A self-gravitating halo consisting of neutrinos
(near the degeneracy limit) can be approximated by a $\gamma=5/3$
polytrope, which has a constant density core with a rapid decline
beyond a core radius $R_c$ (Sanders, 2007). To simplify our
calculations we approximate this polytrope by a $\beta$-model profile
with $\beta=4/3$:

\beq
\rho(r)=\rho_0 \left[1+ \left(\frac{r}{R_c}\right)^2\right]^{-2},
\eeq

\noindent where $\rho_0$ is the central density

\beq
\rho_0=\frac{3 M_\nu}{\pi^2 R_c^3}=\frac{3}{\pi^2 R_c^3}\times 1.4m_\nu M_*,
\eeq

\noindent and the core radius $R_c$, which follows from the polytrope
model, is given by (Sanders 2007)

\begin{equation}
R_c=1.8\left(\frac{m_\nu}{1eV}\right)^{-4/3}
{(\frac{M_\nu}{10^{14}M_{\sun}}\frac{1}{0.06})}^{{1}\over{12}}{~\rm Mpc}.
\end{equation}

We also explored a uniform density sphere with radius $R_c$, and found
that the results are similar to the ones described below.  The
procedure used to compute the corresponding lensing signal in MOND is
outlined in Appendix~A. The (maximum) mass of the neutrino halo is set
by the ratio of the cosmological baryon and neutrino densities, which
for the adopted neutrino mass is approximately three times that of the
stellar mass (e.g., Sanders, 2007). In doing so we assume that only
the three known families of active neutrinos are relevant. A massive
sterile neutrino cannot be ruled out by our analysis. However,
invoking such a particle, which mimicks cold dark matter, would defeat
the rationale behind developing an alternative theory of gravity.

\begin{figure}
\includegraphics[angle=0,width=8.5cm]{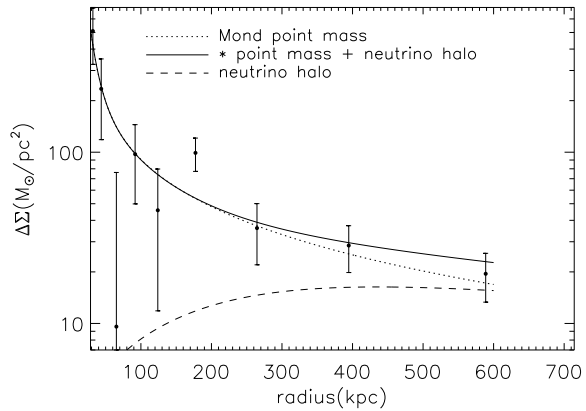}
\caption{Fit to the SDSS data for the highest luminosity bin.  A model
consisting of a point mass of $M_*$ (stellar component) and a neutrino
halo with a mass $3\times M_*$ is fitted to the measurements within
200 kpc. The compact stellar mass dominates the lensing signal 
on small scales and as a result, our results are insensitive to the
effects of a neutrino halo.}
\label{fig:halo}
\end{figure}

Figure~\ref{fig:halo} shows the best fit model to the SDSS
measurements for the highest luminosity bin. Note that the stellar
mass is the only parameter in the fit. The best fit model, with a
stellar mass of $(2.2\pm0.6)\times 10^{12}M_\odot$, yields a core
radius of $R_c=720$ kpc. Hence, the inferred mass-to-light ratio is
$18.3\pm4.7M/L_{r,\odot}$, much higher than what is expected from
realistic stellar populations. Note that the model is fitted to the
measurements within 200 kpc. Figure~\ref{fig:halo} shows that on these
scales the stellar component dominates the lensing signal and that the
neutrino halo only becomes relevant on large scales. This may seem
surprising because the neutrino halo is much more massive than the
stellar component. It is, however, important to recall that the weak
lensing shear measures mass contrasts: for instance, a constant sheet
of matter does not introduce a shear (Gorenstein et al. 1988).  As a
result, it is not the mass of the halo that is most relevant for our
study, but the fact that the halo is very extended and has a large
constant density core.

We verified that our results do not depend our choice of parameters.
First of all, the results are robust against varying the value of
$\beta$. This is not surprising because the size of the constant
density core is the main relevant parameter. To examine possible
(model) uncertainties in the calculation of the core radius, we also
considered an extreme case with $R_c=360$ kpc. We find no qualitative
difference: the core remains too large to produce a sufficiently large
shear at small radii. Finally we note that adopting a different
neutrino mass does not yield satisfactory results. A more massive
neutrino\footnote{Note that a mass larger than 2eV exceeds the
experimental bound from $\beta$-decay} results in a smaller core
radius, but the resulting signal at large radii exceeds the
lensing signal beyond 200kpc. We do not use those scales in our fits,
because they may be biased high. They can, however, be considered
upper limits and models that exceed the observed large scale values
are therefore strongly disfavoured. A lower value for the neutrino
mass results in an even larger core radius {\it and} lower lensing signal.

Consequently, the inferred stellar mass-to-light ratio is reduced only
slightly compared to our original results. This is clearly
demonstrated in the right panel of Figure~\ref{fig:m2l}, which shows
the resulting mass-to-light ratios. We therefore conclude that
including a halo of active neutrinos around the lens galaxies cannot
explain the high mass-to-light ratios. Similar conclusions have also
been reached in studies of galaxy clusters (e.g., Takahashi \& Chiba
2007; Natarajan \& Zhao 2008). Finally we note that these
findings suggest we can safely ignore the contribution from neutrinos
in our study of the scaling relations.

A related issue arises from the fact that the most luminous galaxies
are located in filaments. The filaments can in principle contribute a
significant amount of mass along the line-of-sight, which was studied
in Feix et al. (2007). In their model, the filament induces a shear
which is uniform over scales much larger than the ones we study
here. A constant shear will cancel when we compute the azimuthally
averaged shear around the lenses, and we therefore argue that
filaments do not affect our small scale g-g lensing signal.

Similar to massive clusters of galaxies, many elliptical galaxies are
surrounded by hot, ionized gas. This gas is observed in X-ray
observations (e.g., Humphrey et al., 2006) and we need to consider its
impact on our results. Humphrey et al. (2006) used Chandra
observations to study the ionized gas around NCG720, NGC4125 and
NGC6482, which are three field ellipticals. Hence, these galaxies
resemble the lenses studied in our analysis. In all three cases, the
X-ray observations allow Humphrey et al. (2006) to study the relative
distribution of the stars, gas and dark matter. From their Figure~6 it
is clear that the gas is much more extended than the stellar mass.
Furthermore, the amount of gas inferred from the data is less than the
stellar mass within 200 kpc.

To study the effect of an extended gas distribution we considered a
range of $\beta$ models where we varied to core radius from 1kpc to
30kpc. The mass within 200 kpc was fixed to the stellar mass (as
suggested by the results in Humphrey et al., 2006). The smaller core
radius of the hot gas distribution does boosts the shear on small
scales compared to the neutrino halo, but can only reduce the MOND
mass-to-light ratio values by $\sim 20\%$. 

\section{Conclusions}

We study the amplitude of the weak gravitational lensing signal as a
function of lens luminosity around a sample of relatively isolated
galaxies. We demonstrate how such a study can be used to test Modified
Newtonian Dynamics (Milgrom 1983a). Compared to previous work, our
study is a test of MOND on relatively large scales, where the
differences between MOND and $\Lambda$CDM are expected to be large.
We show that MOND predicts a Tully-Fisher-like relation between the
lensing signal (as quantified by the Einstein radius) and the stellar
mass $M_*$.

Our analysis of data from RCS and SDSS shows that the amplitude of the
lensing signal as a function of stellar mass is well described by a
power law with a best fit slope of $\alpha=0.74\pm0.08$. This result
is inconsistent with $\alpha=0.5$, the predicted slope of
MOND. Uncertainties in the stellar populations used to derive the
stellar masses are expected to be small and should not alter our
result.

As a related test, we determined the MOND mass-to-light ratio as a
function of lens luminosity. Our results require dark matter to
explain the amplitude of the lensing signal for the most luminous
galaxies.  We examined whether our findings can be reconciled with
MOND by considering a massive halo of active neutrinos. Such a halo is
found to be too extended to produce a significant change in the g-g
lensing signal on scales less than 200 kpc. Similarly, hot ionized gas
observed around elliptical galaxies (e.g., Humphrey et al. 2006)
cannot explain our results in the context of MOND.

Although in this paper we focussed on MOND, we note that our findings
are relevant for any alternative theory of gravity. Such theories
should not just attempt to explain the Tully-Fisher relation (i.e.,
the scaling of rotation curves on small scales), but also need to
explain our lensing measurements. In our opinion this is an important
(and non-trivial) test to pass. Similarly, more exotic neutrino
models have been proposed (e.g., Zhao 2008), which can also be tested
through galaxy-galaxy lensing.  Finally we note that much larger
surveys, such as the Canada-France-Hawaii Telescope Legacy Survey
(CFHTLS), will significantly improve the data in the coming years.

\section*{acknowledgments}

It is a pleasure to thank Rachel Mandelbaum for their SDSS data, her
explanation of those data and her comments.  We also thank Jorge
Penarrubia for a careful reading of the manuscript. LT would like to
thank Jiren Liu for useful discussions and Lisa Glass for her help on
the first draft. We also thank the anonymous referee for helping
clarify the paper.

\begin{appendix} 
\section{MOND lensing signal}
\label{mond}

The development of a relativistic form of MOND, known as
Tensor-Vector-Scalar gravity (TeVeS), by Bekenstein(2004) has opened
the way to confront the theory to observations of gravitational
lensing. Fortunately the steps to compute the lensing quantities in
TeVeS the same as those in General Relativity, but with a modified
gravitational potential (Zhao 2006).  In this Appendix we outline the
calculation of the lensing signal in the context of MOND for
(spherically) symmetric cases.

The first step is the calculation of the gravitational potential. In
highly symmetric systems (e.g., spherical halos), the TeVeS potential
can be approximated by a MOND potential (just like GR can be
approximated by Newtonian dynamics).  The MOND gravitational potential
$\phi$ is determined by the equation

\begin{equation}
\label{eqn:mond}
\vec \nabla \cdot [\mu{(|\vec \nabla{\phi}|/a_0)} \vec \nabla{\phi}] = 4 \pi G \rho,
\end{equation}

\noindent where $\rho$ is the mass density and $a_0$ $\approx
10^{-8}$cms$^{-2}$ is the MOND characteristic acceleration. The
function $\mu$ is required to satisfy $\mu(x \gg 1) \approx 1$, so
that Newtonian dynamics is recovered in the limit of large
accelerations and $\mu(x \ll 1) \approx x$.

Note that the Newtonian gravitational potential $\phi_N$ is determined
by the Poisson equation \( \nabla^2{\phi_N} = 4 \pi G \rho \) and that
the Newtonian acceleration is given by \( \vec g_N = - \vec
\nabla{\phi_N} \). Similarly, the MOND acceleration is defined as \(
\vec g = - \vec \nabla{\phi} \). With these definitions, the Newtonian
acceleration $g_N$ and MOND acceleration $g$ are related through a
curl field (Bekenstein \& Milgrom, 1984):

\begin{equation}
\label{eqn:gm1}
\mu{(g/a_0)} \vec g = \vec g_N + \nabla \times \vec h.
\end{equation}

\noindent In highly symmetric systems (i.e. those with spherical,
planar, or cylindrical symmetry), the second curl term in
Eq.(\ref{eqn:gm1}) vanishes (Bekenstein \& Milgrom, 1984) and we have
the exact result

\begin{equation}
\label{eqn:gm2}
\mu{(g/a_0)} \vec g = \vec g_N.
\end{equation}

\noindent Thus, although equation (\ref{eqn:mond}) is non-linear and
is generally difficult to solve, the MOND acceleration $\vec g$ can be
readily obtained from the Newtonian acceleration $\vec g_N$ in
symmetric cases. In this paper we consider spherically symmetric
objects and we adopt $\mu(x)=x/sqrt(1+x^2)$, where $x=g/a_0$. This
yields an acceleration in the MOND regime of $g=\sqrt{a_0 g_N}$.

The next step is the calculation of the lensing signal itself, following
Zhao (2006). In MOND the steps are similar to the GR case, except
that the real mass density $\rho$ is replaced by an effective density
$\rho_{\rm eff}$ which is related to the MOND gravitational potential
by the Poisson equation:

\begin{equation}
\rho_{\rm eff} = \nabla^2{\phi}/4 \pi G.
\end{equation}

We note that $\nabla^2{\phi}$ can be computed using
Eqn.(\ref{eqn:gm2}) and an explicit form of the $\mu$ function,
because $\nabla^2{\phi} = \nabla \cdot (\nabla{\phi}) = \nabla \cdot
(-g)$. Hence, we can obtain the effective density $\rho_{\rm eff}$
through the known Newtonian acceleration $g_N$.

The next step is the calculation of the effective projected surface
density $\Sigma_{\rm eff}$, which is obtained by integrating
the effective density $\rho_{\rm eff}$ along the line of sight:

\begin{equation}
\Sigma_{\rm eff} (x, y) = \int_{-\infty}^{+\infty} \rho_{eff}(x,y,z) \,dz.
\end{equation}

\noindent As discussed in \S2, the convergence $\kappa$ in MOND
is given by

\begin{equation}
\kappa(r) = \Sigma_{\rm eff}(r)/ \Sigma_{\rm crit},
\end{equation}

\noindent where the critical surface density is given by Eqn.~(3).
Finally, we compute the tangential shear using the relation between
$\gamma_T$ and $\kappa$:

\begin{equation}
\label{eqn:tshear}
\gamma_T(r) ={\bar{\kappa}(<r)} - \kappa(r),
\end{equation}

\noindent where $\bar\kappa(<r)$ is the mean convergence within the radius $r$.

\end{appendix}

\label{lastpage}

\end{document}